\documentclass[10pt]{article}
\usepackage[a4paper,textwidth=17.0cm, top=2cm, bottom=3cm]{geometry}%
\usepackage[labelformat=empty,labelsep=quad, skip=0pt]{caption}
\usepackage{graphicx}
\usepackage[superscript,biblabel]{cite}

\usepackage{setspace,caption}
\usepackage{amsmath}
\usepackage{amssymb}
\usepackage{siunitx}
\usepackage{authblk}

\newcommand{\nbar}{\ensuremath{\bar{\mathrm{n}}}}

\newcommand{\tcouple}{\ensuremath{t_\mathrm{C,0-1-2}}}
\newcommand{\tcouplea}{\ensuremath{t_\mathrm{C,0-1}}}
\newcommand{\tcoupleb}{\ensuremath{t_\mathrm{C,2-0}}}

\usepackage{ifthen}

\newcommand{\Omegac}[1][]{%
	\ifthenelse{\equal{#1}{}}{\ensuremath{\Omega_\mathrm{C}}}{\ensuremath{\Omega_\mathrm{C,#1}}}%
}
\newcommand{\omegac}[1][]{%
	\ifthenelse{\equal{#1}{}}{\ensuremath{\omega_\mathrm{C}}}{\ensuremath{\omega_{\mathrm{C,#1}}}}%
}

\newcommand{\Omegacij}[3][]{%
	\ifthenelse{\equal{#1}{}}{\ensuremath{\Omega_{\mathrm{C,}{#2 - #3}}}}{\ensuremath{
			\Omega_{\mathrm{C,#1,}{#2 - #3}}}}%
}

\newcommand{\phieff}{\ensuremath{\Phi_\mathrm{eff}}}

\title{Interference in a Prototype of a two-dimensional Ion Trap Array Quantum Simulator}
\begin{document}

\author[$\dagger$,*]{Frederick~Hakelberg}
\author[$\dagger$,*]{Philip~Kiefer}
\author[$\dagger$]{Matthias~Wittemer}
\author[$\dagger$]{Ulrich~Warring}
\author[$\dagger$]{Tobias~Schaetz}
\affil[$\dagger$]{Albert-Ludwigs-Universität Freiburg, Physikalisches Institut, Hermann-Herder-Strasse 3, Freiburg 79104, Germany.}
\affil[*]{These authors contributed equally to this work.}

\maketitle
{\bfseries
Quantum mechanics dominates various effects in modern research from miniaturizing electronics, up to potentially ruling solid-state physics, quantum chemistry and biology\cite{ciracGoalsOpportunitiesQuantum2012, georgescuQuantumSimulation2014}{}.
To study these effects experimental quantum systems may provide the only effective access\cite{feynmanSimulatingPhysicsComputers1982, haukeCanOneTrust2012}{}.
Seminal progress has been achieved in a variety of physical platforms\cite{georgescuQuantumSimulation2014}, highlighted by recent applications\cite{roushanChiralGroundstateCurrents2017,barredoSyntheticThreedimensionalAtomic2018,zhangObservationManybodyDynamical2017,bernienProbingManybodyDynamics2017}{}.
Atomic ions are known for their unique controllability and are identical by nature, as evidenced, e.g., by performing among the most precise atomic clocks\cite{huntemannSingleIonAtomicClock2016} and providing the basis for one-dimensional simulators\cite{winelandNobelLectureSuperposition2013}{}.
However, controllable, scalable systems of more than one dimension are required to address problems of interest and to reach beyond classical numerics with its powerful approximative methods\cite{ciracGoalsOpportunitiesQuantum2012, haukeCanOneTrust2012}{}.
Here we show, tunable, coherent couplings and interference in a two-dimensional ion microtrap array, completing the toolbox for a reconfigurable quantum simulator.
Previously, couplings\cite{brownCoupledQuantizedMechanical2011, harlanderTrappedionAntennaeTransmission2011} and entangling interactions\cite{wilsonTunableSpinSpin2014} between sites in one-dimensional traps have been realized, while coupling remained elusive in microtrap approaches\cite{kumphOperationPlanarelectrodeIontrap2016, mielenzArraysIndividuallyControlled2016,bruzewiczScalableLoadingTwodimensional2016}{}.
Our architecture is based on well isolatable ions as identical quantum entities hovering above scalable CMOS chips\cite{mielenzArraysIndividuallyControlled2016}. In contrast to other multi-dimensional approaches\cite{brittonEngineeredTwodimensionalIsing2012}{}, it allows individual control in arbitrary, even non-periodic, lattice structures\cite{schmiedOptimalSurfaceElectrodeTrap2009}{}.
Embedded control structures can exploit the long-range Coulomb interaction to configure synthetic, fully connected many-body systems to address multi-dimensional problems\cite{schaetzScalableQuantumSimulations2007}{}.
}\\

Our approach of a synthetic dense lattice (Fig.~1a) is designed for controlling interactions between tens to hundred sites\cite{mielenzArraysIndividuallyControlled2016}{}. To engineer the desired interactions at short and long range within microtrap arrays, we can rely on the established state-dependent interaction with external fields, e.g. with laser fields for the control of effective spin-spin interaction\cite{porrasEffectiveQuantumSpin2004}{}. Thereby, we might address first open problems of interest, such as synthetic gauge fields (Fig.~1b) with phonons tunnelling between sites, accumulating Aharonov-Bohm phases\cite{bermudezSyntheticGaugeFields2011, bermudezPhotonassistedtunnelingToolboxQuantum2012} or accessing quantum spin Hamiltonians (Fig.~1c) using an effective spin-spin interaction to investigate frustration in triangular lattices\cite{schmiedQuantumSimulationHexagonal2011, schneiderExperimentalQuantumSimulations2012}{}.

To introduce the essentials for the inter-site coupling, we consider the simplified case of two ions with equal charge $q$ and mass $m$, trapped in individual harmonic potentials at sites T$_i$, $i=0,1$ (blue and orange in Fig. 1d) with distance $d$. 
At each site we select one motional mode for coupling, with corresponding uncoupled oscillation frequency $\omegac[i]$, phase, and amplitude of coherent motional excitation $\bar n_i$, the local average phonon numbers.
On resonance, $\omegac[0]=\omegac[1]$ and with aligned coupling modes the ions
exchange excitation with the rate %
${\Omegac[res]=\frac{1}{4\pi\epsilon_0}\frac{2 q^2}{d^3 m\omegac[i]}}$
with the vacuum permittivity $\epsilon_0$\cite{brownCoupledQuantizedMechanical2011, harlanderTrappedionAntennaeTransmission2011}{}.
A detuning $\Delta\omegac = \omegac[1] - \omegac[0]\neq0$ increases the coupling rate $\Omegac[det] = \sqrt{\left(\Omegac[res]\right)^2 + \left(\Delta\omegac\right)^2}$ and decreases the exchange efficiency $\kappa_\mathrm{C,det} = \left(\Omegac[res]/\Omegac[det]\right)^2$, 
the fraction of the maximal excitation exchanged between sites. In Fig.~1e we illustrate the time evolution of $\bar n_1$ as a fraction of the total excitation $\bar n_\mathrm{tot} = \sum_i\bar n_i$, initially at T$_0$.
In a higher dimensional arrangement also rotations of the mode orientations are required as tuning parameters:
For motional modes still in a common plane, arbitrary angles $\alpha_0$ and $\alpha_1$ (Fig.~1d) permit tuning the coupling rate $\Omegac[rot] = \Omegac[] \left(\cos\alpha_0 \cos\alpha_1 - 0.5\sin\alpha_0 \sin\alpha_1\right)$, cf. Fig.~1f.
Anharmonic contributions to the trapping potential increase the coupling rate while reducing the coupling efficiency. We combine these effects in an effective coupling rate $\Omegac[eff]$ with corresponding exchange efficiency $\kappa_\mathrm{C,eff}$.

As a building block of complex lattices we realize a triangular array of atomic magnesium ions using a multi-layer surface-electrode trap\cite{seidelinMicrofabricatedSurfaceElectrodeIon2006} sketched in Fig.~1g, microfabricated by Sandia National Labs using scaleable techniques developed for MEMS and CMOS devices\cite{moehringDesignFabricationExperimental2011,schaetzFocusQuantumSimulation2013}{}.
The electrodes are grouped into two radio-frequency (rf) and 30 control electrodes.
We drive the rf electrodes in phase with frequency $\Omega_\mathrm{rf}/(2\pi)\approx\SI{89}{\mega\hertz}$ and amplitude $U_\mathrm{rf}\approx\SI{40}{\volt}$. This provides an effective potential \phieff\ with three trapping sites T$_0$, T$_1$ and T$_2$, approximately \SI{40}{\micro\meter} above the surface at $d\approx\SI{40}{\micro\meter}$, each T$_i$ allowing to trap several ions.
The control electrodes allow for local tuning of an additional control potential $\Phi_\mathrm{c}$ at each T$_i$. We use $\Phi_\mathrm{c}$ to control electric fields (3 degrees of freedom, DOF, per site), and mode orientations and potential curvatures, i.e. $\omegac[i]$ (5 DOF per site), requiring a minimum of $3+5$ electrodes per T$_i$, here 24 in total.
We exploit the current overhead of electrodes to access partial control over the anharmonic contributions of the combined potential $\phieff+\Phi_\mathrm{c}$.
In the experiment, at each site, we use the lowest-frequency mode of the three motional modes as coupling mode with frequency $\omegac[i]/(2\pi)\approx\SIrange{3}{5}{\mega\hertz}$.
We detune the $\omegac[i]$ by $\Delta\omegac[]/(2\pi) \gtrsim \SI{100}{\kilo\hertz}$ mitigating the $\kappa_\mathrm{C,eff}\propto10^{-4}$, to switch off the coupling.
Preparation is performed using global laser beams addressing all T$_i$ for Doppler cooling to thermal states $\bar n_i\lesssim\SI{20}{quanta}$, optionally followed by sideband cooling close to the motional ground state\cite{mielenzArraysIndividuallyControlled2016, kalisMotionalmodeAnalysisTrapped2016}{}. We tune $\Phi_\mathrm{c}$ to ensure similar mode orientations at all T$_i$ with respect to the cooling lasers.
By application of an oscillating voltage to a control electrode, resonant with one $\omegac[i]$ we prepare coherent states of motion at any selected T$_i$ with amplitude of choice of up to several thousand quanta without measurable crosstalk.
Due to the Doppler effect the fluorescence light from the individually illuminated ion is modulated as a non-linear function of $\bar n_i$, permitting to derive motional excitation in the range of $10^2 - 10^4$ quanta and to reconstruct the motional state\cite{kalisMotionalmodeAnalysisTrapped2016}.
For fixed parameters experiments are repeated between 200 and 400 times for averaging and extraction of uncertainties (standard error of the mean, s.e.m.).

To demonstrate the coupling between two sites of the array, we implement the experimental protocol depicted in Fig.~2a:
After global preparation by Doppler cooling we excite the coupling mode at T$_0$ to a coherent state of motion with $\nbar = \SI{2202(57)}{quanta}$.
By real-time shaping of $\Phi_\mathrm{c}$ we tune the coupling modes into resonance in $t_\mathrm{tune}=\SI{10}{\micro\second}$, with a resolution of $2\pi\cdot\SI{0.2}{\kilo\hertz}$ and thereby couple the sites. We choose $t_\mathrm{tune}$ short compared to $(\Omegac[eff])^{-1}>t_\mathrm{tune}$, but long (adiabatic) compared to $(\omegac[i])^{-1}<t_\mathrm{tune}$, thereby ensuring $\bar n_i$ to remain constant. After variable duration \tcouplea\ of up to $\SI{1200}{\micro\second}$ we again decouple the sites.
Finally we locally detect the motional excitation at selected sites. 
The resulting data in Fig. 2b shows the coherent motional exchange between T$_0$ and T$_1$ with a rate of $\Omegac[eff,0-1]/(2\pi)=\SI{1.92(2)}{\kilo\hertz}$.
We confirm the creation of coherent states using calibration measurements including a coherent de-excitation at both sites.
From a sinusoidal model fit, assuming constant total motional excitation and an exponential dephasing term, we derive that $\kappa_\mathrm{C,eff}=\SI{46(2)}{\percent}$ and a dephasing timescale of \SI{800(60)}{\micro\second}. We attribute the $\kappa_\mathrm{C,eff}$ and $\Omegac[eff]$ to the selected anharmonic contributions in the trap. To emphasize the conserved mean total excitation $\nbar_\mathrm{tot}$ we also show residual deviations $\Delta\nbar_\mathrm{tot}$ (Fig. 2b, grey data points).

To demonstrate the real-time control required for sequential coupling of sites in two dimensions we employ the following sequence (Fig. 3a):
We start by Doppler cooling at all T$_i$ and subsequently align the coupling modes at sites T$_0$ and T$_2$. We excite a coherent state of motion with $\bar{n}_{2} = \SI{6880(170)}{quanta}$.
Then T$_0$ and T$_2$ are coupled for a fixed duration $\tcoupleb=\SI{100}{\micro\second}$ transferring $\bar{n}_{0} = \SI{1060(25)}{quanta}$ to T$_0$.
Afterwards we decouple and adiabatically rotate the modes, in \SI{100}{\micro\second} to the initial configuration, aligning the mode orientations at T$_0$ and T$_1$. We then tune the latter sites into resonance for variable \tcouplea. Finally we detect at T$_0$ and T$_1$ individually. The resulting data is shown in Fig. 3b, where we also depict the first transfer of initial motional excitation to T$_0$ during \tcoupleb. The combined sinusoidal model fit yields a coupling rate of $\Omegacij[eff]{0}{1}/(2\pi)=\SI{3.09(6)}{\kilo\hertz}$ where $\kappa_\mathrm{C,eff}=\SI{33(3)}{\percent}$ of the energy is exchanged between T$_0$ and T$_1$, and a dephasing timescale of $\SI{380(35)}{\micro\second}$. We attribute the increased $\Omegacij[eff]{0}{1}$ and decreased $\kappa_\mathrm{C,eff}$ to larger anharmonic contributions compared to the previous measurement.

To demonstrate simultaneous, global coupling and interference of coherent states within the two-dimensional array we present two experiments (Fig. 4a):
After preparation we rotate all three coupling modes towards the centre of the triangle.
In the first realization we excite at T$_2$ and in the second realization at T$_0$ and T$_2$. After both realizations the excitation is followed by simultaneous coupling of all three modes for variable duration \tcouple, and local detection of the motional excitation.
In the first case, Fig.~4b, $\SI{1330(170)}{quanta}$ are transferred simultaneously from T$_2$ to T$_0$ and T$_1$ with $\Omegac[eff]=\SI{2.0(1)}{\kilo\hertz}$ and a dephasing timescale of $\SI{800}{\micro\second}$.
Here the oscillators at T$_0$ and T$_1$ are in phase and exchange among them is suppressed.
We explain the residual difference in $\Omegac[eff]$ by drifts of $\omegac[i]$ in between measurements, and small differences of the individual coupling strengths.
To investigate the case of multiple excitations with different phases, in the second, case we prepare excitation at two sites:
Both T$_2$ and T$_0$ are initialised to approximately \SI{1000}{quanta} (Fig. 4c). The time delay between the excitations of controllable phase relation and the initial detuning lead to a phase shift between the coherent local oscillators that is constant for experimental realizations. The excitation is then exchanged coherently between the three sites.
We describe the time evolution at T$_i$ using sine functions with individual frequencies $\left\{\SI{2.09(8)}{}, \SI{1.80(6)}{}, \SI{2.02(7)}{}\right\}\si{\kilo\hertz}$, peak to peak amplitudes $\left\{\SI{470(59)}{}, \SI{381(33)}{}, \SI{447(40)}{}\right\} \si{quanta}$, and phases.
For this measurement we observe a suppressed dephasing timescale, compared to Fig.~4b, which we partially attribute to the lower excitation amplitudes lowering the effect of the anharmonic contributions.

We realized ground-state cooling in our setup\cite{kalisMotionalmodeAnalysisTrapped2016}{}, while coupling at a further reduced amplitude in the quantum regime is currently limited by motional heating rates on the order of a few $\si{quanta\per\milli\second}$.
Established methods, e.g. surface-cleaning \cite{hite100FoldReductionElectricField2012} can mitigate these heating rates substantially compared to coupling rates.
Additionally, effective spin-spin interactions can be amplified by using more than one ion per site\cite{harlanderTrappedionAntennaeTransmission2011} and by parametrically driving the trapping potential\cite{geTrappedIonQuantum2018}{}.
Realizing the latter by modulating $\Phi_\mathrm{c}$ locally may enable controlling photon-assisted tunnelling\cite{bermudezPhotonassistedtunnelingToolboxQuantum2012} between off-resonant sites of choice.
Further engineering anharmonicities by accessing higher order terms of the trapping potential might allow realizing mode mixing and on-site phonon-phonon coupling.
In upcoming chip traps based on this architecture, tens to hundred sites can be realized\cite{mielenzArraysIndividuallyControlled2016} and inter-site distances might be further reduced to increasing the coupling strength. The trap positions can extend into the third dimension\cite{schmiedOptimalSurfaceElectrodeTrap2009}{}, as already realized for ancilla traps above our array\cite{mielenzArraysIndividuallyControlled2016}{}.
This should allow addressing open problems (Fig.~1) and might assist the development and benchmarking of novel numerical approaches.

\clearpage
%
\begin{figure*}[ht!]
	\centering
	\includegraphics[]{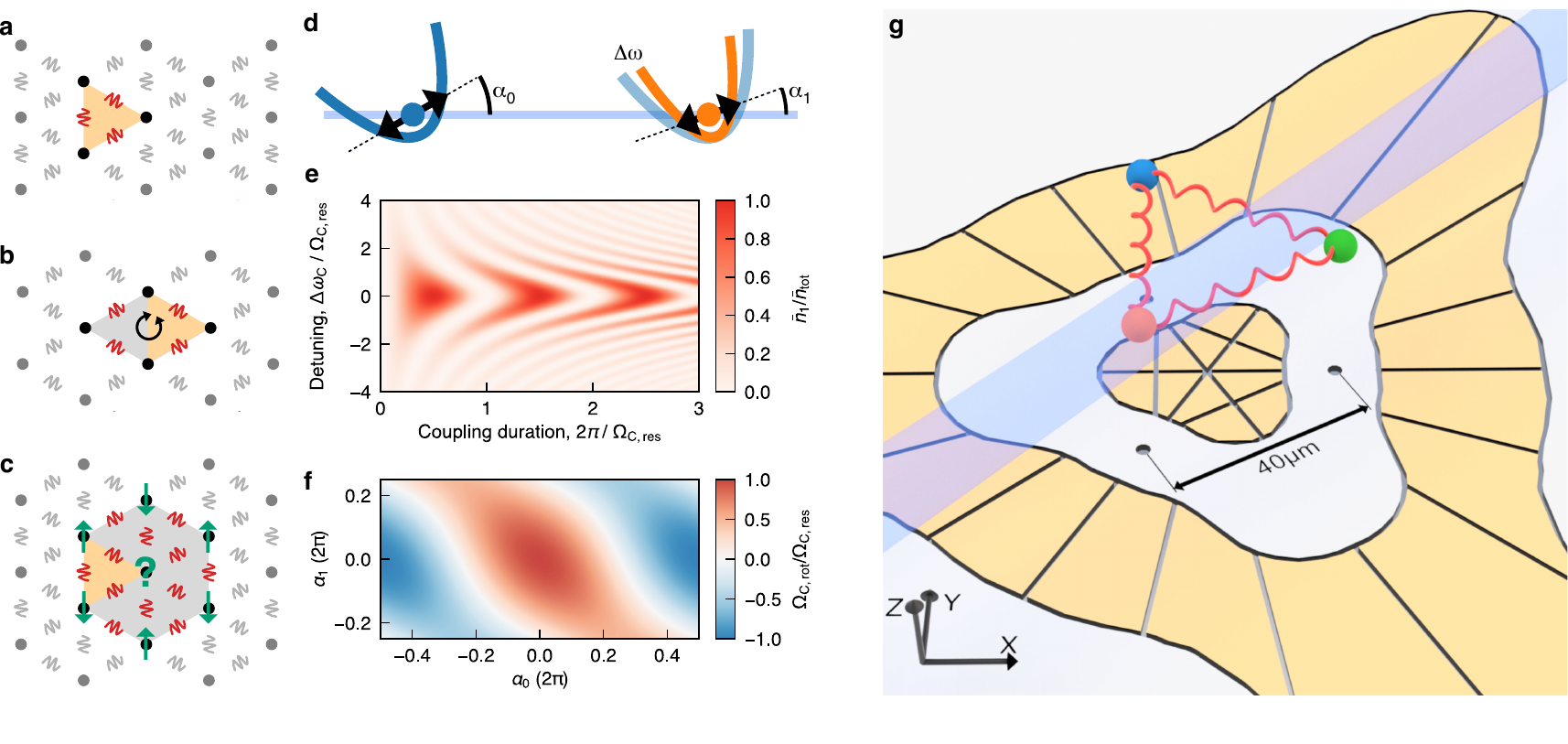}

	\caption{
		\textbf{Fig. 1: Versatile two-dimensional lattice based on individually trapped and locally controlled atomic ions.}
		\textbf{a-c,} Sketch of a dense lattice of ions featuring electronic (effective spin, single-head arrow) and motional (phonon) degrees of freedom constructed from a basic triangle (orange shaded area).
		Mutual Coulomb interaction is used to engineer motional couplings between neighbouring sites (red \& grey springs) as well as at long range.
		This allows realizing experimental quantum simulations; for example, \textbf{b}, to study the emergence of synthetic gauge fields, impinging on ladders of plaquettes, or \textbf{c}, accessing e.g. frustration in quantum spin Hamiltonians.
		\textbf{d,} Basics of tunable inter-site coupling between two ions (blue and orange filled circles) in individual harmonic wells requiring coherent control of local parameters, such as, motional frequencies, mode orientation (double arrows), and excitation.
		\textbf{e,} Tuning the difference of the motional frequencies $\Delta\omegac$ allows adjusting the amount and timescale of the exchange of motional excitation.
		\textbf{f,} Rotating the mode orientation by angles $\alpha_{0}$ and $\alpha_{1}$ (cf.~\textbf{d}) permits full control of sign and amplitude of the coupling strength \Omegac[rot].
		\textbf{g,} Realization of the equilateral triangle of the lattice by scalable CMOS microfabrication technology:
		The surface of the gold-covered microchip contains two radio-frequency (rf in grey) and 30 control electrodes (orange). They provide real-time control for three distinct trapping sites (blue, orange, and green spheres) separated by $\approx\SI{40}{\micro\meter}$. Laser beams (blue) and control electrodes can be used for global and local operations, for example, to prepare, manipulate, couple, entangle, and detect electronic and motional states at each individual site.	
	}
	\label{fig:expsetup}
\end{figure*}

\clearpage
\begin{figure}[ht]
	\centering
	\includegraphics[]{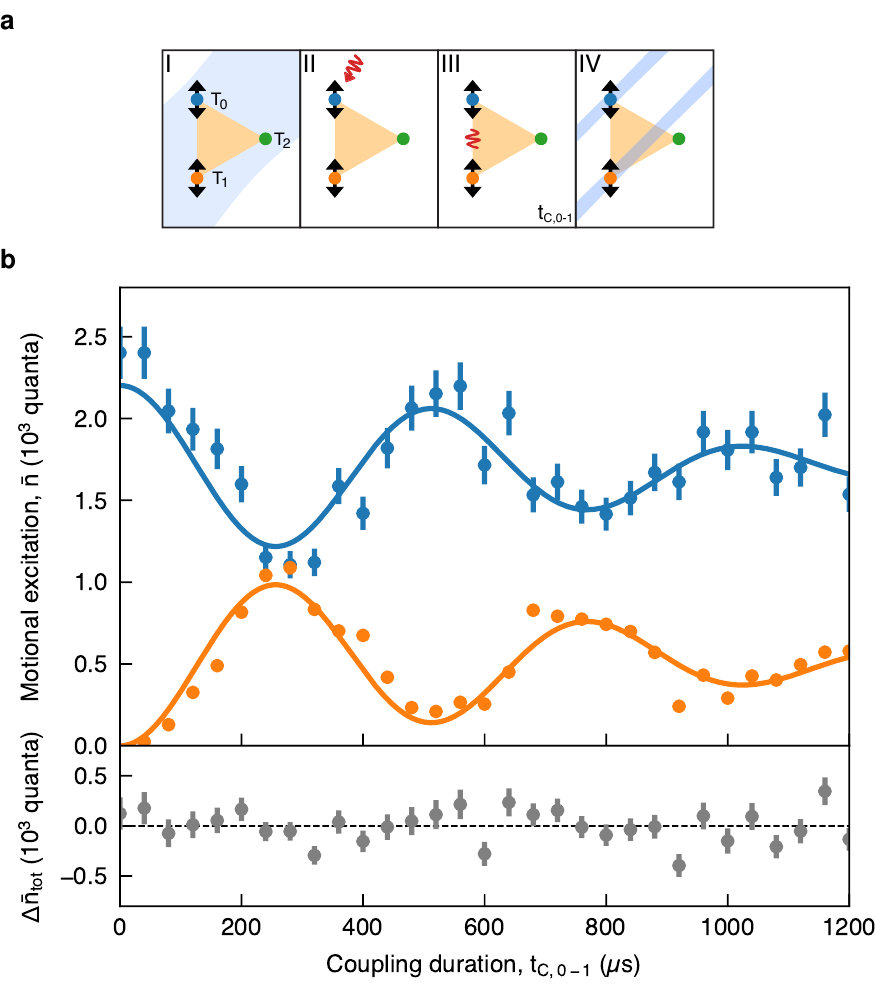}
	
	\caption{
		\textbf{Fig. 2: Coherent motional coupling between two sites.}
		\textbf{a,} Experimental protocol for coupling single ions at sites T$_0$ and T$_1$ (blue and orange filled circles):
		(I) Global preparation via laser cooling (blue shaded area) of all sites.
		(II) Local excitation of the coupling mode at T$_0$ to a coherent state by an oscillating electric field (red wiggle) on resonance.
		(III) Enabling of inter-site coupling by tuning T$_0$ to resonance with T$_1$ for \tcouplea.
		(IV) Local detection of motional excitation using fluorescence induced by focused laser beams.
		\textbf{b,} Observation of coherent exchange of energy between T$_0$ and T$_1$ (blue and orange data points) as a function of coupling duration \tcouplea, errorbars indicate the s.e.m.
		A combined model fit (solid lines, see text) yields exchange with $\Omegac[eff]/(2\pi)= \SI{1.92(2)}{\kilo\hertz}$. Below we show residual deviations from the mean total excitation, $\Delta\nbar_\mathrm{tot}$ (grey data points).	
	}
	\label{fig:fig2}
\end{figure}

\clearpage
\begin{figure}[ht]
	\centering
	\includegraphics[]{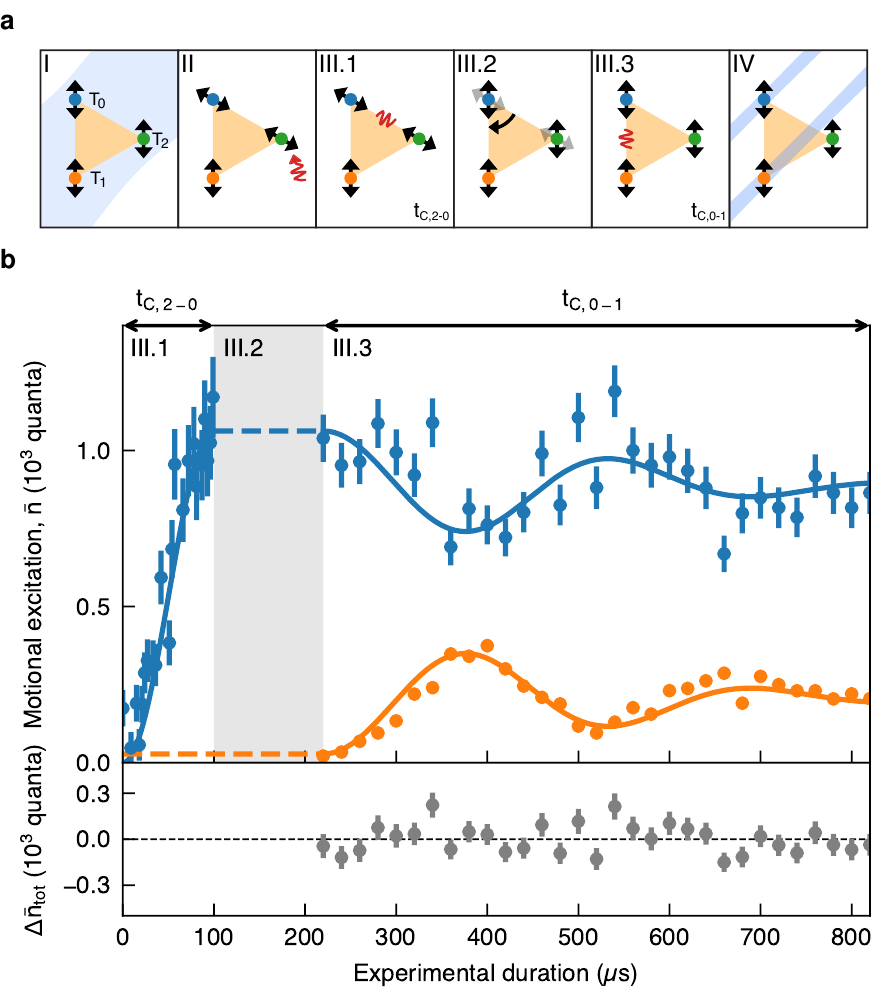}
	
	\caption{
		\textbf{Fig. 3: Real-time sequential coupling of multiple sites.} 
		\textbf{a,} Experimental sequence to coherently couple from T$_2$ via T$_0$ to T$_1$ includes global (I) and local (II) operations, to prepare a coherent motional state in T$_2$.
		(III) Sites are coupled in subsequent steps:
		(III.1) Establishing the first inter-site coupling between T$_2$ and T$_0$ for \tcoupleb.
		(III.2) Decoupling and tuning local electric potentials adiabatically to coherently reconfigure the mode orientation in T$_0$.
		(III.3) Establishing the second coupling to coherently exchange energy between T$_0$ and T$_1$ for variable duration \tcouplea.
		(IV) Tracking evolution of motional quanta via local detection (blue shaded areas) at sites T$_0$ and T$_1$. 
		\textbf{b,} Experimental results showing the dynamics of motional excitation at sites T$_0$ and T$_1$ (blue and orange data points), errorbars indicate the s.e.m. During \tcoupleb\ (III.1) the excitation is transferred from T$_2$ to T$_0$.
		The rotation (III.2) is adiabatic, as indicated by the constant motional excitation (blue dashed line).
		During \tcouplea\ (III.3) excitation is exchanged between T$_0$ and T$_1$ with a rate of $\Omegac[eff,0-1]/(2\pi)=\SI{3.09(6)}{\kilo\hertz}$ (solid lines, see text). Below we show residual deviations from the mean total excitation, $\Delta\nbar_\mathrm{tot}$ (grey data points).	
	}
	\label{fig:fig3}
\end{figure}

\clearpage
\begin{figure}[ht!]
	\centering
	\includegraphics[]{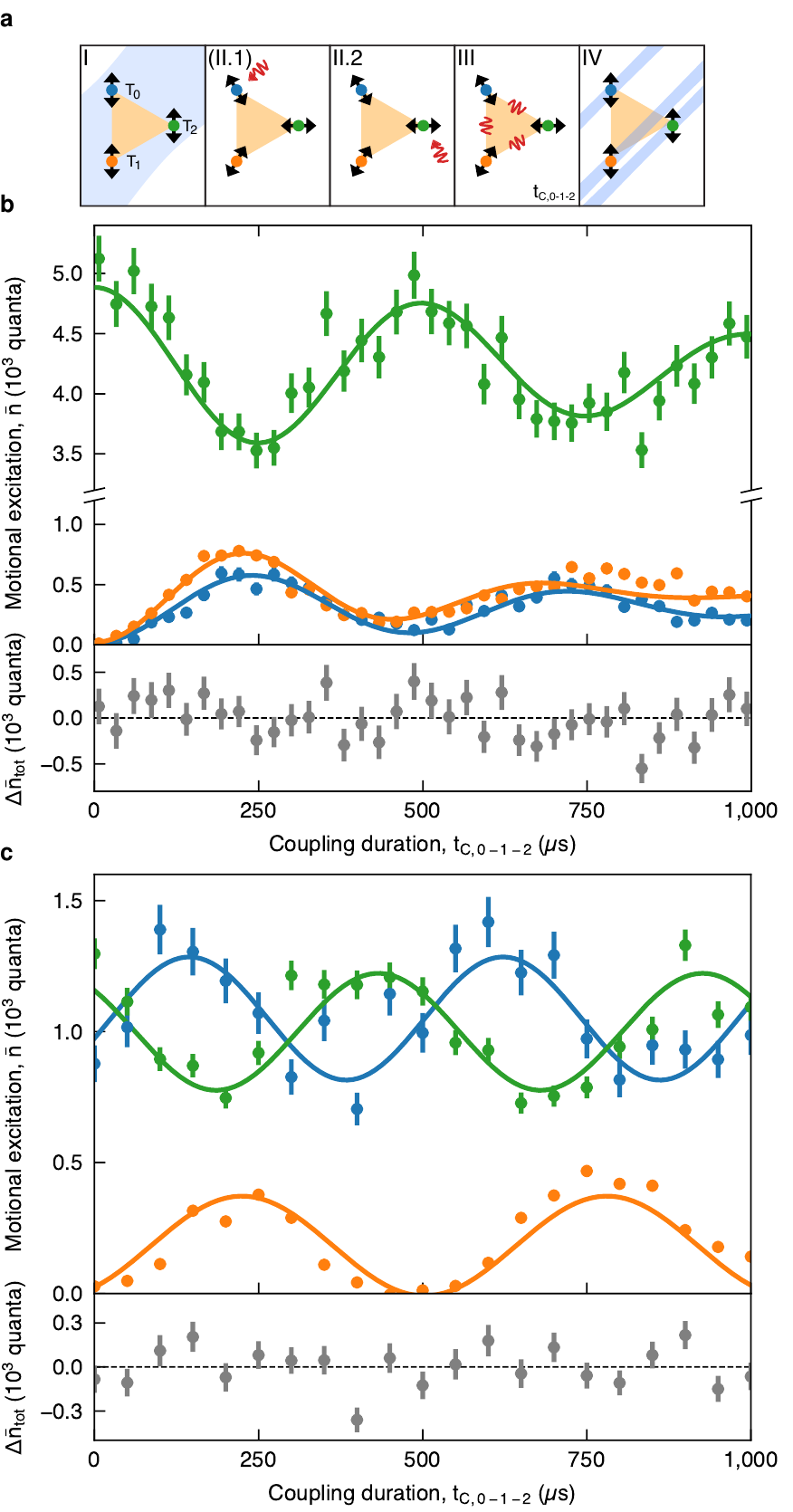}
	
	\caption{
		\textbf{Fig. 4: Simultaneous coupling of all sites and interference.}
		\textbf{a,} The experimental sequence comprises:
		(I) Global state initialisation.
		(II) Rotation of all coupling modes towards the centre of the array and optional excitation of a coherent state of motion, either (II.2) at T$_2$ only, or (II.1 + II.2) at T$_0$ and subsequently at T$_2$.
		(III) Coupling of all three sites for variable duration \tcouple.
		(IV) Local detection of all sites (blue shaded areas).
		\textbf{b,} Omitting (II.1) only excitation at T$_2$ (green data points) is simultaneously transferred to T$_0$ and T$_1$ (blue and orange data points), and back to T$_2$. Errorbars indicate the s.e.m.
		\textbf{c,} In (II.1) and (II.2) we excite both T$_2$ and T$_0$ to approximately \SI{1000}{quanta} with a fixed phase shift between the coherent local oscillators. These phase relations rule the coherent exchange between the three sites and yield individual frequencies $\approx\SI{2}{\kilo\hertz}$, amplitudes, and phases. To emphasize the conserved mean total excitation we show residual deviations $\Delta\nbar_\mathrm{tot}$ below (grey data points). The fixed phase relation allows to observe the interference of the coherent states of motion exchanging between the three sites.
	}
	\label{fig:fig4}
\end{figure}

\clearpage

\section*{Acknowledgements}
We thank J.-P. Schr\"oder for help with the experimental control system. The trap chip was designed in collaboration with R. Schmied in a cooperation with the NIST ion storage group and produced by Sandia National Laboratories. We thank A. Bermudez and D. Porras for fruitful discussions. This work was supported by the Deutsche Forschungsgemeinschaft (DFG) [SCHA 973/6-3].

\section*{Author Contributions}
F.H., P.K. participated in the design of the experiment, built the experimental apparatus, collected data and analysed the results. M.W., U.W. and T.S. participated in design and analysis of the experiment. F.H. and T.S. wrote the manuscript. All authors discussed the results and the text of the manuscript.

\section*{Author Information}
The authors declare no competing financial interest.

\section*{Data Availability}
The datasets generated during and/or analysed during the current study are available from the corresponding author on reasonable request.
\end{document}